\documentclass[prd,twocolumn,superscriptaddress,floatfix,nopacs,preprintnumbers,nofootinbib,10pt,aps]{revtex4-2}
\usepackage[utf8]{inputenc}

\usepackage[caption = false]{subfig}
\usepackage[normalem]{ulem}
\usepackage{xcolor}
\usepackage{mathtools}

\usepackage{footnote}
\usepackage{mathrsfs}
\usepackage{physics}
\usepackage{amssymb,bm}
\usepackage{amsmath}
\usepackage{mathtools}
\usepackage{slashed}
\usepackage{graphics}
\usepackage{graphicx}
\usepackage{tikz}
\usepackage{comment}
\usepackage{physics}

\newcommand{\nc}{N_\mathrm{c}}

\newcommand{\gev}{\mathrm{GeV}}
\newcommand{\jpsi}{$\mathrm{J}/\psi$ }
\newcommand{\bt}{\boldsymbol{b}_\perp}
\newcommand{\rt}{\boldsymbol{r}_\perp}

\newcommand{\xt}{\boldsymbol{x}_\perp}

\newcommand{\lqcd}{\Lambda_\mathrm{QCD}}
\newcommand{\yt}{\boldsymbol{y}_\perp}
\newcommand{\Deltat}{\boldsymbol{\Delta}_\perp}
\newcommand{\as}{\alpha_\mathrm{s}}

\usepackage[breaklinks,colorlinks,citecolor=citcolor,urlcolor=blue,linkcolor=lcolor]{hyperref}

\definecolor{lcolor}{rgb}{0.5,0,0}
\definecolor{citcolor}{rgb}{0,0.3,0.0}

\begin{document}

\author{Patricia Gimeno-Estivill}
\email{patricia.p.gimenoestivill@jyu.fi}
\author{Tuomas Lappi}
\email{tuomas.v.v.lappi@jyu.fi}
\author{Heikki Mäntysaari}
\email{heikki.mantysaari@jyu.fi}
\affiliation{
Department of Physics, University of Jyväskylä,  P.O. Box 35, 40014 University of Jyväskylä, Finland
}
\affiliation{
Helsinki Institute of Physics, P.O. Box 64, 00014 University of Helsinki, Finland
}

\title{Inclusive \texorpdfstring{$\mathrm{J}/\psi$}{J/psi} production in forward proton-proton and proton-lead collisions at high energy}

\begin{abstract}
We calculate the cross section for forward $\mathrm{J}/\psi$ production in proton-proton and minimum bias proton-lead collisions using the Color Glass Condensate (CGC) and Non-Relativistic QCD (NRQCD) formalism consistently with Deep Inelastic Scattering data. The cross section for color singlet states is sensitive to a 4-point correlator of Wilson lines for which we describe in the Gaussian approximation in the large-$N_c$ limit. Furthermore, we quantify the importance of finite-$\nc$ corrections to have a small $\sim 12\%$ effect. In contrast with the generic NRQCD expectation, we show that the color singlet contribution is only $15\%$ to the total cross section.  We also compare our predictions for $\mathrm{J}/\psi$ production as well as for the nuclear modification ratio $R_{pPb}$ to LHCb and ALICE data. We find a good agreement at forward rapidities, except at low transverse momenta where the cross sections are overestimated.   
\end{abstract}

\maketitle 

\section{Introduction}
Abundant data on \jpsi production in proton-proton and proton-nucleus collisions at high-energy is available from different measurements performed at RHIC \cite{STAR:2018smh,PHENIX:2019brm,PHENIX:2019ihw} and at the LHC \cite{LHCb:2015foc,LHCb:2017ygo,ALICE:2013snh,ALICE:2021dtt, ALICE:2022zig,ALICE:2018mml}. These processes are especially powerful in probing QCD dynamics in the region where gluon saturation effects are expected to be important. This is because the \jpsi mass is optimally suitable for such studies: it is heavy enough to ensure the validity of  perturbative calculations, but low enough for saturation effects to be clearly visible.
Furthermore, because of the heavy quark mass, it should also be possible to understand the quark-antiquark hadronization process in a weak coupling framework.

In this paper, we consider \jpsi production at forward rapidities, where a proton-nucleus collision can be described in the Color Glass Condensate (CGC) effective field theory as an interaction between a ``dilute'' proton probe and a ``dense'' system of gluons in the target nucleus~\cite{Kovchegov:2012mbw,Gelis:2010nm,Weigert:2005us,Iancu:2003xm}. This picture is valid at high energy (small momentum fraction $x$), where gluon saturation is predicted by the CGC. In this framework, the lowest-order (LO) process in the strong coupling constant $\as$ is a projectile gluon splitting into a charm
quark pair either before or after interacting with the target. The production amplitude for a heavy quark pair at LO in $\as$ was computed in Ref.~\cite{Blaizot:2004wv} in the dilute-dense approximation.

Historically~\cite{Chang:1979nn, Baier:1981uk} it was argued that the dominant contribution to \jpsi production at small momentum transfer $p_\perp$ comes from the fusion of two gluons in a color singlet channel. This  color singlet model (CSM)
is still widely used in the literature, see e.g. Ref.~\cite{Lansberg:2019adr} for a review. In more recent studies \cite{Ma:2016exq,Cheung:2018tvq,Fujii:2013gxa,Ducloue:2015gfa}, the color octet contributions were also included in the Color Evaporation Model (CEM), which results in a good agreement with RHIC and LHC data.

Nevertheless, the overall normalization in  the  CEM model is typically adjusted by hand to measured cross section data, limiting the universality and predictive power of the approach. This calls for a more systematical framework, which can be provided  by the Non-Relativistic QCD (NRQCD) formalism~\cite{Bodwin:1994jh}. Here the starting point is the observation that the heavy mass of the quarks makes the quarkonium system approximatively non-relativistic. As a consequence, its hadronization into a physical \jpsi particle can then be factorized and described in terms of  long-distance matrix elements (LDMEs)~\cite{Kramer:2001hh}. These are universal non-perturbative quantities that can be extracted from experimental data~\cite{Brambilla:2010cs}.  

Our goal is to test the predicted contribution of the different intermediate quantum states $\kappa$ to the \jpsi production cross section in forward proton-proton (p+p) and minimum bias proton-lead (p+Pb) collisions in the CGC+NRQCD framework. 
In this NRQCD factorization formalism, the \jpsi production cross section for different $c\bar c$ quantum states was first calculated in Ref.~\cite{Kang:2013hta} and phenomenological applications were reported in Refs.~\cite{Ma:2014mri,Ma:2015sia}. In the latter, an approximated form of the 4-point quadrupole correlator of Wilson lines describing the interaction of the probe was considered. In contrast to these works, we use the quadrupole derived in Ref.~\cite{Dominguez:2011wm} in the large-$\nc$ limit in the Gaussian approximation, and we also quantify the importance of the finite-$\nc$ corrections. Unlike earlier phenomenological studies using CGC+NRQCD, the quadrupole and dipole correlators present in the cross section are evolved with the Balitsky-Kovchegov (BK) equation~\cite{Balitsky:1995ub,Kovchegov:1999yj} from an initial condition with all the  parameters obtained consistently from Deep Inelastic Scattering (DIS) data.
We also use a realistic description of the nuclear geometry with an impact parameter ($\bt$)-dependent saturation scale following Ref.~\cite{Lappi:2013zma}, an approach that has been successfully used in Refs.~\cite{Mantysaari:2019nnt,Ducloue:2016pqr,Ducloue:2015gfa,Mantysaari:2023vfh}. Our results for the differential cross section in proton-proton (p+p) and proton-lead (p+Pb) collisions as well as the nuclear modification factor are compared to the LHCb~\cite{LHCb:2015foc,LHCb:2017ygo} and ALICE data~\cite{ALICE:2018mml}. 

We begin this paper with a description of the CGC+NRQCD framework in Sec.~\ref{NRQCD factorization formalism}. The phenomeonological study is presented in   Sec.~\ref{Results}, where we first quantify the importance of finite-$\nc$ corrections to the quadrupole correlator, and then compare our predictions for the \jpsi spectra and the nuclear modification factor to the LHCb and ALICE data. We conclude with a summary in Sec.~\ref{Conclusions}.
\section{CGC+NRQCD factorization formalism}
\label{NRQCD factorization formalism}
The cross section for \jpsi production with transverse momentum $\boldsymbol{p}_\perp$ and rapidity $y$ is expressed in NRQCD factorization
as~\cite{Kang:2013hta}
\begin{equation}
    \frac{\dd\sigma^{\mathrm{J}/\psi}}{\dd^2\boldsymbol{p}_\perp dy}=\sum_\kappa \frac{\dd\hat{\sigma}^{\kappa}}{\dd^2\boldsymbol{p}_\perp \dd y}\langle \mathcal{O}_{\kappa}^{J/\psi}\rangle \;.
    \label{eq:cross section}
\end{equation}
Here, the short distance partonic cross sections $\dd\hat{\sigma}^\kappa$ can be calculated perturbatively in the CGC framework and represent the production of a intermediate heavy quark pair in a state $\kappa$. The sum in Eq.~\eqref{eq:cross section} includes all the possible quantum configurations $\kappa=\prescript{2s+1}{}{L}^{[c]}_J$, where $S,L$ and $J$ are respectively the spin, orbital angular momentum and total angular momentum. The superscript $c$ denotes the color state of the intermediate quark pair, which can be in a color singlet $(c =1)$ or octet $(c =8)$ configuration. On the other hand, the low-energy hadronization of these intermediate heavy quark states into a physical bound state is represented by the long-distance matrix elements (LDME) $\langle\mathcal{O}_\kappa^{\mathrm{J}/\psi}\rangle$. They are universal and non-perturbative, determined by fitting experimental data~\cite{Brambilla:2010cs}.

\subsection{Quarkonium production: short-distance partonic cross sections}
\label{sec:quarkonium production}
\label{Quarkonium production: short distance hard partonic cross sections}

\begin{figure}[tbh]
    \centering
    \includegraphics[width=1\linewidth]{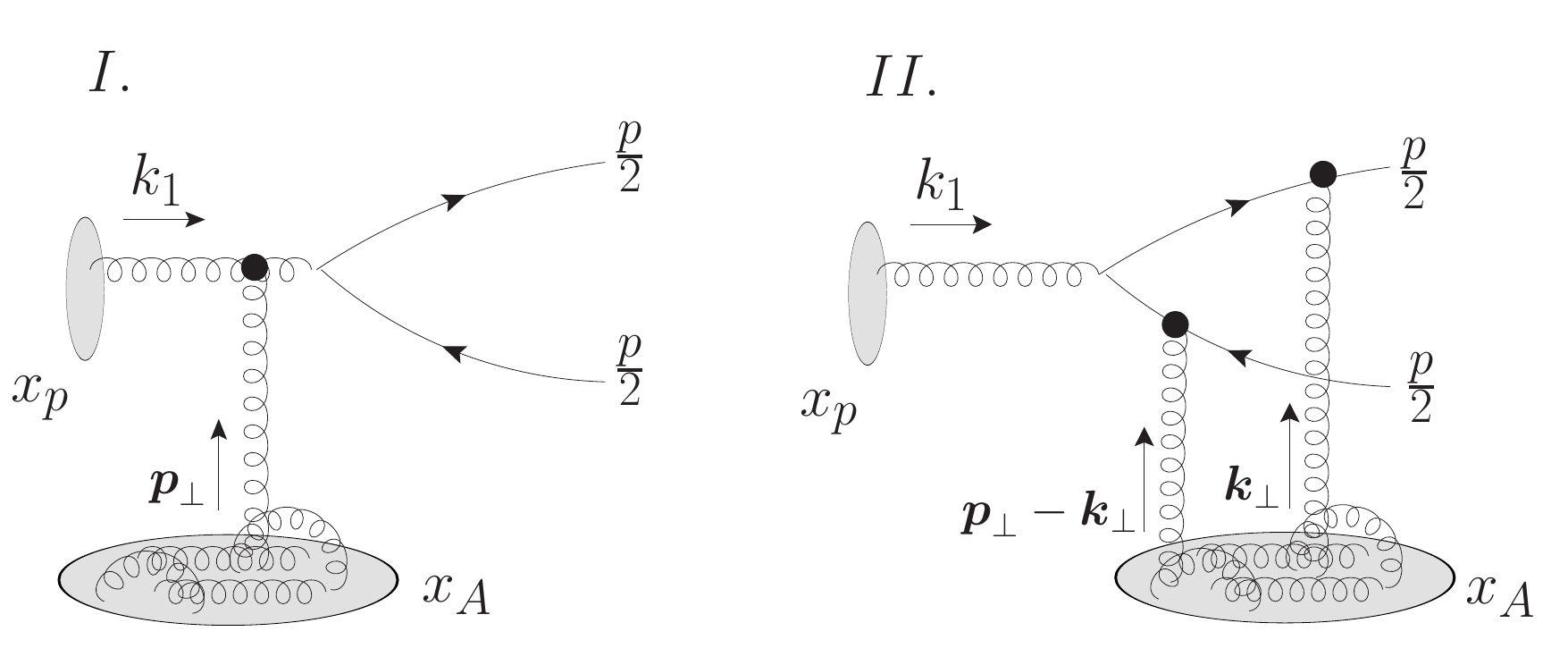}
    \caption{Proton-nucleus collision where a gluon with momentum $k_{1}$ from the projectile proton splits into an on-shell quark and antiquark with momentum $p/2$ either after (diagram I) or before (diagram II) interacting with the gluon field in the target nucleus. The black dots denote the Wilson Lines that resum the multiple interactions of the partons when crossing the shockwave. 
    The complete calculation of the amplitude for this process can be found in Ref.~\cite{Blaizot:2004wv}.}
    \label{Fig: quarkonium production}
\end{figure}

The short-distance creation of a heavy quark pair in forward proton-nucleus collisions is depicted in Fig.~\ref{Fig: quarkonium production}. There, the symbol $x_p (x_A)$ is the fraction of the proton (nucleus) longitudinal momentum carried by the produced particle,
\begin{equation}
    x_{p,A}=\frac{\sqrt{m_{\mathrm{J}/\psi}^2+\boldsymbol{p}_\perp^2}}{\sqrt{s}}e^{\pm y} \;,
\end{equation}
where $\sqrt{s}$ is the center-of-mass energy, $y=\mathrm{ln}(p^+/p^-)/2$ the rapidity and $\boldsymbol{p}_\perp$ the transverse momentum of the produced \jpsi particle with mass $m_{\mathrm{J}/\psi}=3.097$~GeV $(\approx2m_c)$. We use the light-cone notation $x^{\pm}=(x^0\pm x^3)/\sqrt{2}$ for coordinates and $p^+\approx \sqrt{2}E \gg p^-=(m_{\mathrm{J}/\psi}^2+\boldsymbol{p}_\perp^2)/2p^+$ for momenta. 
The large rapidity $y$ in the forward region implies that $x_A \ll 1$ and $x_p \sim 1$, so that the collision can be described as a dilute probe interacting with the small-$x$ gluon field of the target nucleus in the CGC formalism. Further, we consider the proton in the collinear limit where the incoming parton has no transverse momentum ($k_{1\perp}\rightarrow0$). This is because the transverse momentum of the gluons in the proton is much smaller than the mass and transverse momentum of the produced \jpsi particle: $ k_{1\perp} \sim \lqcd \ll m_{\mathrm{J}/\psi}$ and $k_{1\perp} \sim \lqcd \ll p_{\perp}$. This collinear gluon can split into a quark-antiquark pair before or after the collision where transverse momentum from the small-$x_A$ gluons in the nucleus is transferred to the partons (gluon or quark pair).

In this framework, the short-distance partonic cross sections for the heavy quark pair production in color octet (CO) or color singlet (CS) states respectively read~\cite{Kang:2013hta},
\begin{multline}
        \frac{\dd\hat{\sigma}}{\dd^2\boldsymbol{p}_\perp \dd y}\stackrel{\text{CO}}{=}\int\limits_{\bt}\frac{\as}{4(2\pi)^3(N_c^2-1)}x_pf_{p/g}(x_p,\mu^2)\\
        \cross \int\limits_{\boldsymbol{k}_\perp}\mathcal{N}(\boldsymbol{k}_\perp) \mathcal{N}(\boldsymbol{p}_\perp-\boldsymbol{k}_\perp)\tilde{\Gamma}_8^\kappa(p_\perp,k_\perp) \;,
    \label{cross section octet}
\end{multline}
and
\begin{multline}
        \frac{\dd\hat{\sigma}}{\dd^2\boldsymbol{p}_\perp \dd y}\stackrel{\text{CS}}{=}\int\limits_{\bt}\frac{\as}{4(2\pi)^3(N_c^2-1)}x_pf_{p/g}(x_p,\mu^2)\\
        \cross\int\limits_{\boldsymbol{\Delta}_{\perp},\boldsymbol{r}_\perp,\boldsymbol{r'}_\perp}e^{-i\boldsymbol{p}_\perp\boldsymbol{\Delta}_\perp}\Big( Q - D_{\boldsymbol{r}_\perp} D_{\boldsymbol{r}'_\perp}\Big)\tilde{\Gamma}_1^\kappa(r_\perp,r'_\perp) \;,
    \label{cross section singlet}    
\end{multline}
where $Q := Q_{\frac{\rt}{2},\Deltat+\frac{\rt'}{2},\Deltat-\frac{\rt'}{2},-\frac{\rt}{2}}$ is the quadrupole, and we use the short-hand notation $\int_\perp=\int \dd[2]_\perp$. For proton-proton collisions, we assume that the impact parameter dependence factorizes and the integral over the impact parameter $\bt$ results in the transverse area of the proton. We use the value $\pi R^2_{p}=16.36$~mb, which is determined in the fit to DIS structure function data~\cite{Lappi:2013zma} that we use in this work. On the other hand, when we consider minimum bias proton-nucleus collisions,  we account for the impact-parameter profile of the nucleus explicitly. Here $x_pf_{p,g}(x_p,\mu^2)$ is the proton collinear gluon distribution function, for which we use the MSHT20 LO pdf~\cite{Bailey:2020ooq} with $\mu^2=m_{\mathrm{J}/\psi}^2$. The sensitivity of our results on the scale choice is quantified in Sec.~\ref{sec:results}. The strong coupling is set to $\as=0.24$~\cite{Bodwin:1994jh} and it is a source of uncertainty in the normalization of the cross section.  

The hard matrix elements $\tilde{\Gamma}_{8}^\kappa(p_\perp,k_\perp)$ and $\tilde{\Gamma}_{1}^\kappa(r_\perp,r'_\perp)$ are given in appendices B.1 and B.2 in Ref.~\cite{Kang:2013hta}. Moreover, $\boldsymbol{r}_\perp=\boldsymbol{x}_\perp - \boldsymbol{y}_\perp$ is the transverse size of the dipole shown in Fig.~\ref{Fig: quarkonium production}, and $\boldsymbol{r}'_\perp$ is the corresponding size in the complex conjugate amplitude. The coordinate $\Deltat=(\xt'+\yt'-\xt-\yt)/2$ represents the distance between the centers of the dipoles in the amplitude and in the complex conjugate amplitude.

The functions $\mathcal{N}(\boldsymbol{k}_\perp)$ and $ \mathcal{N}(\boldsymbol{p}_\perp-\boldsymbol{k}_\perp)$ in the cross section for color octet states in Eq.~(\ref{cross section octet}) are the Fourier transformed 2-point correlators (dipoles) in momentum space,
\begin{equation}
    \mathcal{N}(\boldsymbol{k}_\perp)=\int\limits_{\boldsymbol{r}_\perp}e^{i\boldsymbol{k}_\perp\boldsymbol{r}_\perp}D_{\boldsymbol{r}_\perp} \;,
\end{equation}
where 
\begin{equation}
    D_{\boldsymbol{r}_\perp }=\frac{1}{\nc}\big\langle \mathrm{Tr}\big[ V_F(0)V_F^\dagger(\boldsymbol{r}_\perp)]\big\rangle_{y_A} \;.
    \label{eq: dipole correlator}
\end{equation}
This dipole is the eikonal (high-energy) scattering amplitude of a quark-antiquark pair off a dense target gluon field, where $\langle\dots\rangle_{y_A}$ represents the average over the color charge densities of the target nucleus evaluated at the evolution rapidity $y_A = \ln\frac{x_0}{x}$~\cite{Kovchegov:2012mbw}. The Wilson Lines $V_{F,A}(\boldsymbol{x}_\perp)$ are defined as    
\begin{equation}
    V_{F,A}(\boldsymbol{x}_\perp)=\mathcal{P}\mathrm{exp}\Bigg(-ig\int \dd x^+ A^-_a(x^+,\boldsymbol{x}_\perp) t^a\Bigg) \;,
    \label{eq: WL}
\end{equation}
where $\mathcal{P}$ is a path ordering operator and $A^-$ the classical gluon field in the target nucleus. The $\mathrm{SU}(3)$ color matrices $t^a$ are in the fundamental (F) or adjoint (A) representation for  quark and gluon probes, respectively. 

The energy (Bjorken-$x$) dependence of the dipole correlator in Eq.~\eqref{eq: dipole correlator} can be obtained by solving the running coupling Balitsky-Kovchegov (rcBK) equation~\cite{Balitsky:1995ub,Kovchegov:1999yj,Balitsky:2006wa} in the large-$N_c$ limit. 
In Ref.~\cite{Lappi:2013zma} the initial condition for this evolution at $x_0=0.01$ was considered to take a form inspired by the McLerran-Venugopalan (MV) model as~\cite{McLerran:1993ni} 
\begin{equation}
\label{eq: MV model}
    \mathcal{N}(\boldsymbol{r}_\perp)=1-\mathrm{exp} \Bigg[-\frac{(\boldsymbol{r}_\perp^2 Q^2_{s0})^\gamma}{4}\mathrm{ln}\Bigg(\frac{1}{|\boldsymbol{r}_\perp|\Lambda_\mathrm{QCD}}+e_c\cdot e \Bigg) \Bigg] \;,
\end{equation}
where $\Lambda_\mathrm{QCD}=0.241$ GeV.  We use here the MV$^e$ fit of Ref.~\cite{Lappi:2013zma}, where  $\gamma$ is fixed to  $\gamma=1$ and the free parameters $Q_{s0}^2=0.060~\mathrm{GeV^2}$ and $e_c=18.9$ are determined by a fit to the HERA inclusive DIS cross section data~\cite{H1:2009pze}. For a nuclear target, the rcBK evolution equation  is solved separately for each impact parameter $\bt$ using an initial condition where the $\bt$-dependence of the saturation scale is obtained from the optical Glauber model:
\begin{multline}\label{eq:Aic}
    \mathcal{N}^A(\boldsymbol{r}_\perp,\boldsymbol{b}_\perp)=1-\mathrm{exp}{\Bigg[ -AT_A(\boldsymbol{b}_\perp) \pi R_p^2 \frac{(\boldsymbol{r}_\perp^2Q_{s0}^2)^\gamma}{4} \Bigg]}\\
    \cross \mathrm{ln}\Bigg(\frac{1}{|\boldsymbol{r}_\perp|\Lambda_{QCD}}+e_c\cdot e \Bigg) \;.
\end{multline}
Here $T_A(\boldsymbol{b}_\perp)$ is the Woods-Saxon nuclear density normalized to unity: $\int \dd[2]{\bt} T_A(\bt)=1$. In the region where the nuclear saturation scale would fall below that of the proton, we replace the dipole-nucleus scattering amplitude by the dipole-proton amplitude scaled such that all non-trivial nuclear effects vanish. Further details can be found in Ref.~\cite{Lappi:2013zma}.
We emphasize that in Eq.~\eqref{eq:Aic} there are no additional free parameters for the nucleus apart from the standard Woods-Saxon density. Thus, the nuclear modification of cross sections is a pure prediction of the framework without any additional fit parameters.

In Eq.~\eqref{cross section singlet}, the cross section for color singlet states is sensitive to dipoles $(D)$ and a quadrupole $(Q)$. The quadrupole is defined as
\begin{equation}
\begin{aligned}
     Q=\frac{1}{N_c}\left\langle \mathrm{Tr}\big[ V_F(\boldsymbol{x}_\perp)V_F^\dagger(\boldsymbol{x}'_\perp)V_F(\boldsymbol{y}_\perp)V_F^\dagger(\boldsymbol{y}'_\perp)]\right\rangle_{y_A} \;,
    \label{eq: quadrupole}
\end{aligned}
\end{equation}
where $Q:=Q_{\xt\xt'\yt\yt'}$. This expression is not reducible to a product of dipoles neither for a large nucleus nor in the large-$N_c$ limit~\cite{Dumitru:2010ak}.
However, in the Gaussian approximation \cite{Marquet:2010cf,Fujii:2006ab} the color charge correlators are assumed to be Gaussian  even after evolution and the quadrupole can be rewritten in terms of two-point correlation functions.
Unlike previous studies of inclusive heavy quarkonium production, we use the explicit Gaussian  approximation for the quadrupole
in the large-$N_c$ limit as well as at finite $N_c$.
The explicit expressions are given respectively by Eq.~(B.22) and Eq.~(B.21) in Ref.~\cite{Dominguez:2011wm}. 
The accuracy of the Gaussian approximation at the initial condition of the small-$x$ JIMWLK evolution~\cite{Mueller:2001uk} and after a relatively long evolution has been shown in Ref.~\cite{Dumitru:2011vk}. The importance of an accurate description of the quadrupole in the case of inclusive dihadron production has been demonstrated in Ref.~\cite{Lappi:2012nh}.
\subsection{Hadronization: long-distance matrix elements} \label{Hadronization: Long-Distance Matrix Elements}

In the NRQCD formalism, the \jpsi state  can be written as a power series in the relative velocity $v\ll1$ of the heavy quark pair~\cite{Cho:1995vh},
\begin{equation}
\label{eq:VectorExpansion}
\begin{aligned}
    |\mathrm{J}/\psi\rangle=&O(1)|\prescript{3}{}{S}^{[1]}_1\rangle+O(v)|\prescript{3}{}{P}^{[8]}_Jg\rangle+ O(v^{2})|\prescript{1}{}{S}^{[8]}_0g
    \rangle\\ &+O(v^2)|\prescript{3}{}{S}^{[8]}_1gg\rangle+\dots
\end{aligned}
\end{equation}
where the dominant Fock state involves the quark-antiquark pair in a color singlet state and quantum numbers that are consistent with the physical vector meson\footnote{Quarkonium  $\prescript{2S+1}{}{L}^{[c]}_J$ states have parity $P=(-1)^{L+1}$ and charge conjugation $C=(-1)^{L+S}$.} with $J^{PC}=1^{--}$. The quantum numbers of higher Fock states are selected according to conservation of total angular momentum $J$, parity $P$ and charge conjugation $C$ in the heavy quark Lagrangian and their contribution in the velocity expansion in Eq. (\ref{eq:VectorExpansion}) is determined by the energy shift with respect to the dominant state~\cite{Kramer:2001hh}. The soft gluons present in higher Fock states can be emitted before the bound state is formed and thus changing the color and spin of the intermediate heavy quark pair. This effect is included in universal long-distance matrix elements (LDMEs) which can be written as vacuum expectation values of 4-fermion operators with the NRQCD heavy quark Lagrangian \cite{Bodwin:1994jh}. Their hierarchy in powers of velocity is established according to NRQCD power counting rules \cite{Bodwin:1994jh} and their specific values are determined by fitting decay width or cross sections data. The four independent LDMEs used in the phenomenology study of \jpsi production in Sec.~\ref{Results} are listed in Table~\ref{table: LDME}.

\begin{table}[hbt!]
\renewcommand{\arraystretch}{1.5}
\begin{tabular}{lc}
\hline \hline
$\langle \mathcal{O}(\prescript{3}{}{S}^{[1]}_1)\rangle$ \:  $O(v^0)$& $1.16/(2\nc) \: \mathrm{GeV}^3$ \\
\hline 
$\langle \mathcal{O}(\prescript{1}{}{S}^{[8]}_0)\rangle $ \:  $O(v^3)$ & $0.089 \pm 0.0098 \: \mathrm{GeV}^3$ \\
\hline
$\frac{\langle \mathcal{O}(\prescript{3}{}{P}^{[8]}_0)\rangle}{m_c^2}$    \;\;  $O(v^4)$ & $0.0056 \pm 0.0021 \: \mathrm{GeV}^3$ \\
\hline
$\langle \mathcal{O}(\prescript{3}{}{S}^{[8]}_1)\rangle$ \:  $O(v^4)$ & $0.0030 \pm 0.0012 \: \mathrm{GeV}^3$  \\
\hline\hline
\end{tabular}
\caption{
Values of LDMEs used in this work and their corresponding order in velocity.
The LDME for the color singlet state $\prescript{3}{}{S}^{[1]}_1$ is estimated using the value of the \jpsi wavefunction at the origin in the Buchmüller and Tye potential model in Ref.~\cite{Eichten:1995ch}. The LDMEs for the color octet states $\prescript{1}{}{S}^{[8]}_0$,$\prescript{3}{}{P}^{[8]}_0$ and $\prescript{3}{}{S}^{[8]}_1$ are determined by fitting NLO collinearly factorized pQCD+NRQCD results to Tevatron prompt \jpsi yields in Ref.~\cite{Chao:2012iv}. The statistical uncertainties correspond to the mass dependence of the LDMEs.}
\label{table: LDME}
\end{table}
The $P$-wave LDMEs with $J=1,2$ that contribute at the same order in velocity as the operator $\langle \mathcal{O}(\prescript{3}{}{P}^{[8]}_0)\rangle$ can be obtained using heavy quark spin symmetry \cite{Bodwin:1994jh}:
\begin{equation}
\begin{aligned}
    \langle \mathcal{O}(\prescript{3}{}{P}^{[8]}_J)\rangle&=(2J+1)\langle \mathcal{O}(\prescript{3}{}{P}^{[8]}_0)\rangle[1+\mathcal{O}(v^2)] \;.
\end{aligned}
\end{equation}
Accordingly, the contribution of the P-wave states in the cross section is 
\begin{equation}
\begin{aligned}
    &\sum_{J=0}^2 \langle \prescript{3}{}{P}^{[8]}_J \rangle \dd\hat{\sigma}^{\prescript{3}{}{P}^{[8]}_J}\\ 
    &=\langle \prescript{3}{}{P}^{[8]}_0 \rangle \dd\hat{\sigma}^{\prescript{3}{}{P}^{[8]}_0} + 3\langle \prescript{3}{}{P}^{[8]}_0 \rangle \dd\hat{\sigma}^{\prescript{3}{}{P}^{[8]}_1} +5\langle \prescript{3}{}{P}^{[8]}_0 \rangle \dd\hat{\sigma}^{\prescript{3}{}{P}^{[8]}_2} \\
    &= 9 \langle \prescript{3}{}{P}^{[8]}_0 \rangle \dd\hat{\sigma}^{\prescript{3}{}{P}^{[8]}_{\mathrm{avg}}} \;, 
\end{aligned}
\end{equation}
where we have defined the following weighted average P-wave short-distance coefficient:    
\begin{equation}
     \dd\hat{\sigma}^{\prescript{3}{}{P}^{[8]}_{\mathrm{avg}}}=\frac{1}{9}\Big(\dd\hat{\sigma}^{\prescript{3}{}{P}^{[8]}_0}+3\dd\hat{\sigma}^{\prescript{3}{}{P}^{[8]}_1}+5\dd\hat{\sigma}^{\prescript{3}{}{P}^{[8]}_2}\Big) \;.
\end{equation}

Although the color singlet LDME is dominant in powers of velocity, when the hard matrix elements $\tilde{\Gamma}_{8}^\kappa(p_\perp,k_\perp)$ and $\tilde{\Gamma}_{1}^\kappa(r_\perp,r'_\perp)$  in Eqs.~(\ref{cross section octet}) and (\ref{cross section singlet}) are expanded\footnote{The correlators are assumed to not contribute any power behavior. We refer the reader to Ref.~\cite{Kang:2013hta} for details of the calculation.} in powers of $m_{\mathrm{J}/\psi}/p_\perp$, the color octet channels become important at large transverse momentum $p_\perp \sim Q_s \gg m_{\mathrm{J}/\psi}$. In this region, by normalizing the $\prescript{3}{}{S}^{[8]}_1$ channel as $\mathcal{O}(1)$, the other color octet channels behave as $\mathcal{O}(m_{\mathrm{J}/\psi}^2/p_\perp^2)$ and the color singlet $\prescript{3}{}{S}^{[1]}_1$contribution is suppressed by $\mathcal{O}(m_{\mathrm{J}/\psi}^4/p_\perp^4)$~\cite{Kang:2013hta}. Therefore, the expectation from a pure power counting argument would be to have  the color singlet state  dominant at small $p_\perp \ll M$, accompanied by a higher contribution from color octet channels at high $p_\perp$\footnote{In the proton collinear approximation, all the quarkonium transverse momentum is given by the nuclear target at small-$x$, with  parametrically $k_\perp \sim Q_s$. Hence, the kinematic region $p_\perp\gg Q_s$ would also get a larger contribution from next-to-leading order (NLO) processes in $\alpha_s$ where a recoiling particle can balance the quarkonium momentum $p_\perp$ in the final state.}.

\section{Results}
\label{sec:results}

Given the dipole-proton scattering amplitude obtained from DIS fits~\cite{Lappi:2013zma} and the values of the LDMEs in Tab.~\ref{table: LDME}, we can next calculate cross sections for inclusive forward \jpsi production in proton-proton (p+p) and proton-lead (p+Pb) collisions. We emphasize that our results are genuine predictions with no free parameters: the dipole amplitude has been fit to DIS data, the extension to nuclei only requires a standard Wood-Saxon density, and the LDMEs have been extracted in a different kinematical regime completely independently from our analysis.

\label{Results}
\subsection{Finite-$\nc$ corrections}

The quadrupole operator in Eq.~(\ref{eq: quadrupole}) cannot be factorized in terms of dipoles. Instead, as already discussed in Sec.~\ref{sec:quarkonium production}, we evaluate it in the Gaussian approximation for which the finite-$\nc$ corrections can be included following Ref.~\cite{Dominguez:2011wm}. As an input, the Gaussian approximation requires a dipole amplitude $D_{\rt}$ obtained by solving the BK equation in the large-$\nc$ limit. 

Although parametrically one expects $\sim 10\%$ finite-$\nc$ corrections to the BK evolution, such corrections in practice turn out to be much smaller, of the order $0.1\%$ at leading order~\cite{Kovchegov:2008mk} (and somewhat larger at next-to-leading order~\cite{Lappi:2020srm}). Hence, when we determine the importance of the finite-$\nc$ corrections for \jpsi production, it is enough to include finite-$\nc$ terms in the quadrupole operator.

\begin{figure}[hbt!]
    \centering
    \includegraphics[width=1\linewidth]{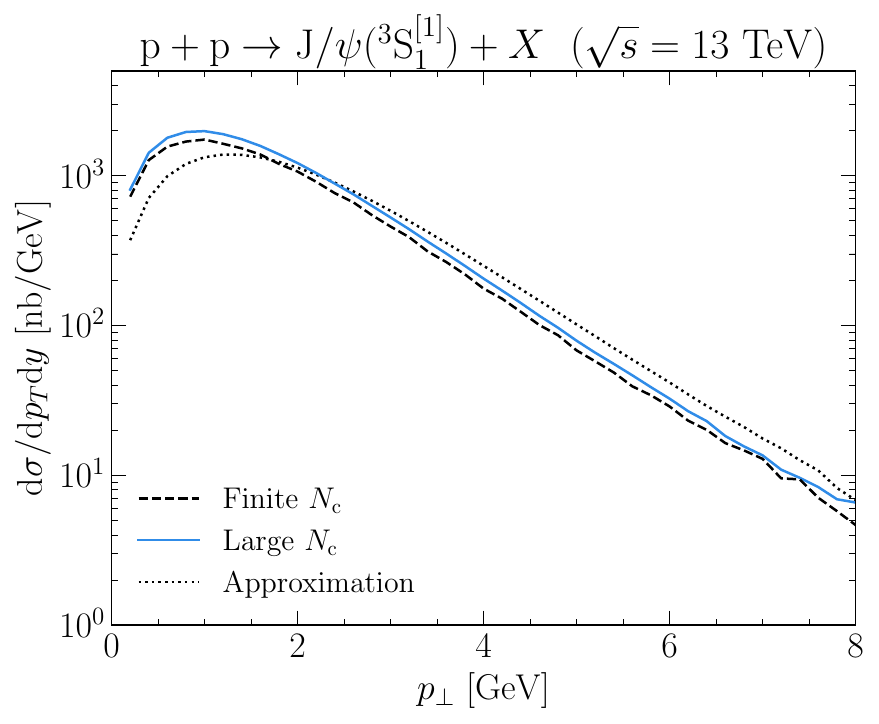}
    \caption{Differential \jpsi production cross section through the color singlet channel $\prescript{3}{}{S}^{[1]}_1$ in proton-proton collisions at centre-of-mass energy $\sqrt{s}=13$ GeV and rapidity $y=3.25$. The dashed line is calculated with the complete expression of the quadrupole at finite-$N_c$, the solid line corresponds to the calculation with the quadrupole in the large-$N_c$ limit and the dotted line is computed with the quadrupole approximation taken from Ref.~\cite{Ma:2014mri}. }
    \label{fig: quadrupole comparison}
\end{figure}

Using the Gaussian approximation, we show in Fig.~\ref{fig: quadrupole comparison} the $p_\perp$ spectra for the inclusive \jpsi production cross section in p+p collisions in the color singlet channel, where the \jpsi is in the ${}^3S_1^{[1]}$ state\footnote{Note that the color octet channel does not depend on the quadrupole and does not have a similar finite-$\nc$ correction.}. The finite-$\nc$ correction is found to be small, of the order of $1/N_c^2\sim12\%$ at all $p_\perp$. Thus, we consider the large-$\nc$ limit to be a suitable approximation in the leading order calculation presented in this work, although finite-$\nc$ corrections might be required for precision level data comparisons.

Additionally, we show in Fig.~\ref{fig: quadrupole comparison} the cross section obtained by using an approximative form of the quadrupole, written in Eq.~(5) of Ref.~\cite{Ma:2014mri} and used in previous phenomenological studies~\cite{Ma:2015sia,Stebel:2021bbn,Ma:2018qvc}. The approximative quadrupole is found to underestimate the cross section at low $p_\perp$ and overestimate it at high-$p_\perp$, maximally by a factor $\sim 2$.  Based on this result, we conclude that a more accurate description of the quadrupole operator e.g. in the Gaussian approximation employed here is necessary in order to accurately predict the cross section in the color singlet channel.

\subsection{CGC+NRQCD in small collision systems}
\begin{figure}[hbt!]
    \centering
    \includegraphics[width=1\linewidth]{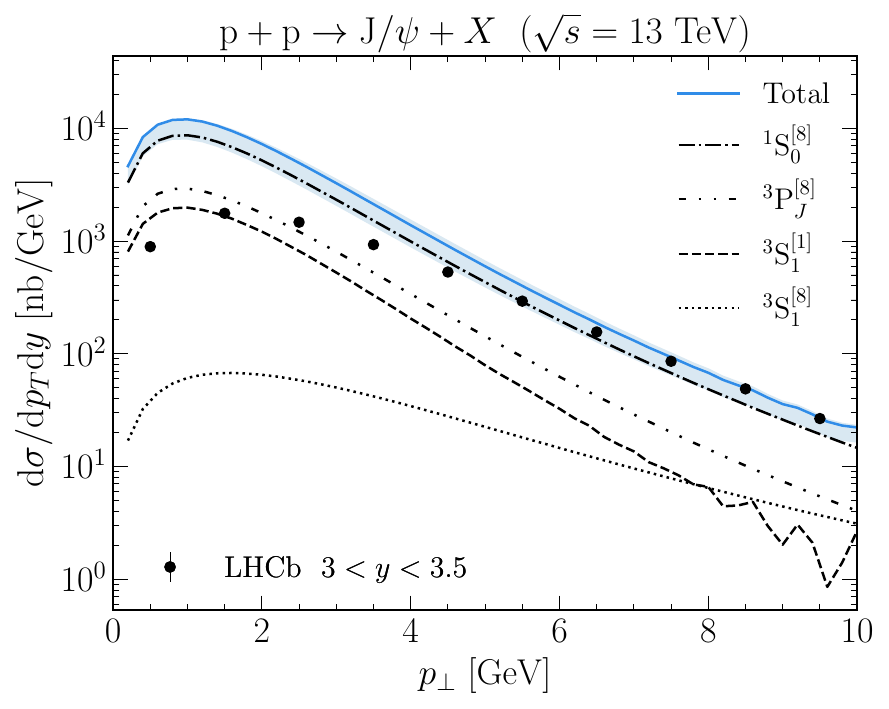}
   \caption{Differential \jpsi production cross section in proton-proton collisions as a function of $p_\perp$, rapidity $y=3.25$ and centre of mass energy $\sqrt{s}=13$ TeV. The sum of the different channels is represented by the top solid line in blue, which is followed by $\prescript{1}{}{S}^{[8]}_0$, $\prescript{3}{}{P}^{[8]}_J$,$\prescript{3}{}{S}^{[1]}_1$ and $\prescript{3}{}{S}^{[8]}_1$ channels, from top to bottom. The factorization scale is fixed to $\mu^{2}=m_{\mathrm{J}/\psi}^{2}$ and the blue uncertainty band includes the variation from $\mu^{2}=m_{c}^{2}$ to $\mu^{2}=m_{\mathrm{J}/\psi}^{2}+p_{\perp}^{2}$. The experimental LHCb data for prompt \jpsi production is taken from Ref.~\cite{LHCb:2015foc}.}
    \label{fig: pp cross section}
\end{figure}

\begin{figure}[hbt!]
    \centering
    \includegraphics[width=1\linewidth]{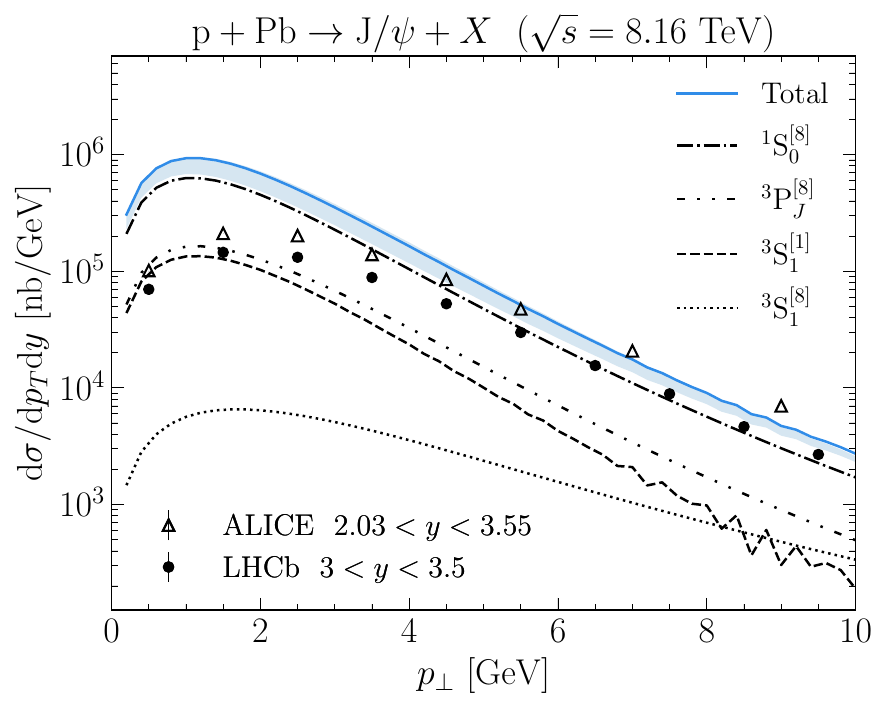}
    \caption{Differential \jpsi production cross section in proton-lead collisions as a function of $p_\perp$, fixed rapidity $y=3.25$ and energy $\sqrt{s}=8.16$ TeV. The sum of the different channels is represented by the top solid line in blue, which is followed by $\prescript{1}{}{S}^{[8]}_0$, $\prescript{3}{}{P}^{[8]}_J$,$\prescript{3}{}{S}^{[1]}_1$ and $\prescript{3}{}{S}^{[8]}_1$ channels, from top to bottom. The factorization scale is fixed to $\mu^{2}=m_{\mathrm{J}/\psi}^{2}$ and the blue uncertainty band includes the variation from $\mu^{2}=m_{c}^{2}$ to $\mu^{2}=m_{\mathrm{J}/\psi}^{2}+p_{\perp}^{2}$. The LHCb data for prompt \jpsi production is taken from Ref.~\cite{LHCb:2017ygo} and the ALICE data for inclusive \jpsi production from Ref.~\cite{ALICE:2018mml}.}
    \label{fig: pA cross section}
\end{figure}

In Figs. \ref{fig: pp cross section} and \ref{fig: pA cross section} we show the transverse momentum distribution of the produced \jpsi in  p+p and p+Pb collisions, respectively. The results are compared with the LHCb~\cite{LHCb:2015foc,LHCb:2017ygo} and ALICE~\cite{ALICE:2018mml} data. The ALICE data corresponds to inclusive \jpsi production and thus also includes a non-prompt contribution not included in our calculation. As it is illustrated in both figures, the differential cross section as a sum of the individual channels overestimates the experimental data by a factor $\sim 5\dots 10$ at small $p_\perp$. Nevertheless, a good agreement of the data is found at moderately large $p_\perp\gtrsim 5$ GeV.

Looking at the individual contributions in Figs.~\ref{fig: pp cross section} and~\ref{fig: pA cross section}, the dominant channel throughout the $p_\perp$ spectra is the color octet $\prescript{1}{}{S}^{[8]}_0$ channel. This has a contribution of about $70\%$ on the total $p_\perp$ integrated cross section in both p+p and p+Pb collisions. Although  NRQCD  predicts the color singlet LDME as the largest in powers of velocity, the relative contribution of the $\prescript{3}{}{S}^{[1]}_1$ channel is approximately $15\%$ of the total in both collisions. By itself, the color singlet channel is close to the experimental data at low $p_\perp$, but always below the color octet contribution. It also falls faster than the color octet channels at high $p_\perp$,  in accordance with the expected momentum dependence discussed in the previous subsection \ref{Hadronization: Long-Distance Matrix Elements}. 

We attribute the behavior of the $p_\perp$ spectra to the fixed leading order calculation in $\alpha_s$ and expect that the implementation of the Sudakov term in the cross section is needed to correct the peak position at small $p_\perp$~\cite{Watanabe:2015yca}. 
This factor should be different for the color singlet and octet channels and thus also affect their relative contributions at a specific $p_\perp$. In particular, the color octet should have a larger Sudakov factor than the singlet, which would increase the mean $p_\perp$ in the octet channels more. Thus, we expect that the inclusion of Sudakov effects would lead to a larger supression of the color octet channels at small $p_\perp$ than the singlet. A refinement of the cross section including this soft gluon radiation could be done in a future work, e.g. among the lines proposed in Ref.~\cite{Sun:2012vc}.

In order to determine the sensitivity on the factorization scale $\mu^2$ we also show in Figs.~\ref{fig: pp cross section} and \ref{fig: pA cross section} the dependence of the total cross section on the scale choice $\mu^2$. The central values are obtained using our default scale choice $\mu^2=m_{\mathrm{J}/\psi}^2$, and we vary the scale from $\mu^2=m_c^2$ to $\mu^2=m_{\mathrm{J}/\psi}^2+p_T^2$. The dependence on the scale is modest, affecting the overall normalization up to $\sim 28\%$, and in particular it can not explain the fact that the presented calculation overestimates the ALICE and LHCb at low $p_T$.

Inclusive \jpsi production in the color evaporation model (CEM) considers a similar description for the target but the hadronization is described with a common transition probability, leading to a dominance of  color octet channels.
The $p_\perp$ spectra obtained in Refs.~\cite{Ducloue:2015gfa,Ducloue:2016pqr} are very similar to our current calculation, which uses the same dipole amplitude, and similarly steeper than in the LHCb data. On the other hand, the CGC+NRQCD formalism has been used to successfully reproduce the p+p and p+Pb data in Refs.~\cite{Ma:2014mri,Ma:2015sia}, although the contribution of different production channels to the total cross section was not shown in the figures. However our setup includes various improvements that explain the different cross sections obtained. First, in Refs.~\cite{Ma:2014mri,Ma:2015sia} the approximative form of the quadrupole illustrated in Fig.~\ref{fig: quadrupole comparison} was used, along with an impact parameter-independent description of the nucleus and a different choice for the factorization scale $\mu^2$. Furthermore the proton transverse area $\pi R_p^2$, that controls the overall normalization, was not taken from the applied DIS fit.
Also, an additional $30\%$ systematic uncertainty from higher order $\as$ corrections was added on top of the statistical uncertainties from the LDMEs. This led to a rather wide error band on the results.

We  also note that the LDME are not known very precisely, for example, there are sets of LDMEs extracted at next-to-leading order (NLO) with negative values \cite{Gong:2012ug,Bodwin:2015iua,Butenschoen:2011ks} and therefore they could lead to negative cross sections. Furthermore, the initial condition for the small-$x$ BK evolution determined from HERA data including only light quarks has its own uncertainties that could affect the cross section, especially at high $p_T$~\cite{Casuga:2023dcf}. Leading order fits have not been able to simultaneously describe the total cross section and the charm production data, and  only light quarks are included in the fit of Ref.~\cite{Lappi:2013zma} that we use here. Consequently the heavy quark contribution is encoded in the non-perturbative fit parameters, and when the fit is used to calculate charm production one typically expects to overestimate the cross section, which is also the case here. 

A simultaneous description of the total cross section and heavy quark production HERA data  is only obtained at NLO accuracy~\cite{Hanninen:2022gje}. One could speculate that, similarly, an NLO calculation is needed to correctly describe the normalization of inclusive \jpsi  cross sections in hadronic collisions. In particular, the Sudakov effect originated from soft gluon radiation \cite{Mueller:2012uf} would be needed to improve our calculation. Here our focus has been to present a self-consistent LO calculation without any free parameters, which we believe is the correct starting point for including higher order effects in a systematic way in the future.
\subsection{Nuclear modification factor and Cronin effect}

As discussed in the previous section, there are several uncertainties in the overall normalization of the cross section. However, many of these cancel in the cross section ratio. Thus we calculate the nuclear modification factor $R_{pPb}$, for which our approach provides a robust parameter-free prediction. 
For minimum bias collisions $R_{pPb}$ is defined as:
\begin{equation}
    R_{pPb}=\frac{\frac{{\dd\sigma_{pPb}}}{\dd^2\boldsymbol{p}_\perp \dd y}}{A\times \frac{\dd\sigma_{pp}}{\dd^2\boldsymbol{p}_\perp \dd y}} \;.
    \label{eq: modifRatio}
\end{equation}


\begin{figure}[tb!]
    \centering
    \includegraphics[width=1\linewidth]{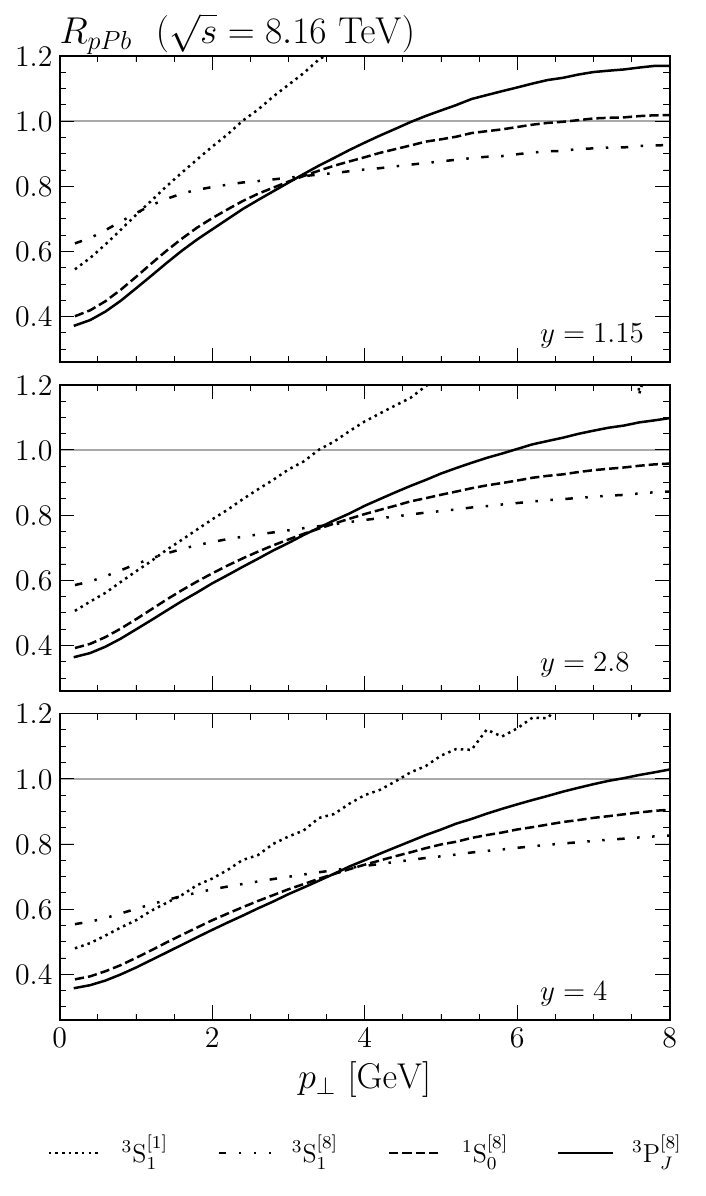}
    \caption{Nuclear suppression factor $R_{pPb}$ for each independent channel as a function of $p_\perp$ and fixed rapidity. The Cronin effect is reduced as the rapidity increases from $y=1.5$ to $y=4$.}
    \label{fig: R_rapidities}
\end{figure}

Here $A=208$ for the lead nucleus. In Fig.~\ref{fig: R_rapidities} we show  $R_{pPb}$ for each independent channel at different rapidities.
We predict a strong nuclear suppression, $R_{pPb}\sim 0.4$, at low \jpsi transverse momentum, resulting from saturation effects.
In agreement with the results presented in Ref.~\cite{Ma:2015sia}, there is a strong Cronin enhancement in the color singlet $\prescript{3}{}{S}^{[1]}_1$ channel at intermediate $p_T$ due to the requirement of an additional gluon exchanged from the target to form a color singlet quark pair. However, as shown in Ref.~\cite{Albacete:2003iq}, the Cronin effect is reduced with BK evolution when the rapidity increases from $y=1.15$ to $y=4$. At large $p_\perp$, the ratio  $R_{pPb}$ approaches unity  by construction~\cite{Lappi:2013zma}.

\begin{figure}[tb!]
    \centering
    \includegraphics[width=1\linewidth]{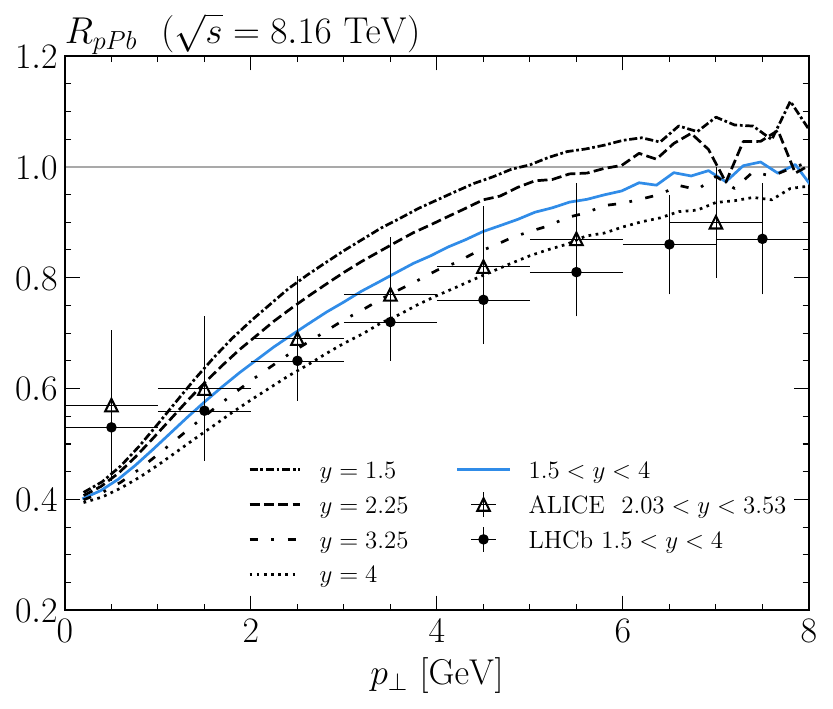}
    \caption{Nuclear supression factor $R_{pPb}$ as a function of $p_\perp$ at $\sqrt{s}=8.16$ TeV for individual rapidities $y=1.5,2.25,3.25$ and $4$ as well as for the integrated value over $1.5<y<4$. The LHCb data~\cite{LHCb:2017ygo} for prompt \jpsi production and ALICE data~\cite{ALICE:2018mml} for inclusive \jpsi production include statistical and systematic uncertainties as quadratic sums in the vertical lines. The horizontal error bars are bin widths.}
    \label{fig: R(pt)}
\end{figure}

The nuclear modification factor for \jpsi production as a function of transverse momentum including all production channels is shown in Fig.~\ref{fig: R(pt)}.  The level of $R_{pPb}$ reproduces well the rapidity integrated data in the $1.5<y<4$ range, although the $p_\perp$-dependence in our calculation is slightly steeper. The reason for this could be the missing next-to-leading order effects, which are expected to lead to a weaker $p_\perp$ dependence due to the additional phase space available for the emission of an extra parton~\cite{Ducloue:2017kkq,Mantysaari:2023vfh}. 

\begin{figure}[tb!]
    \centering
    \includegraphics[width=1\linewidth]{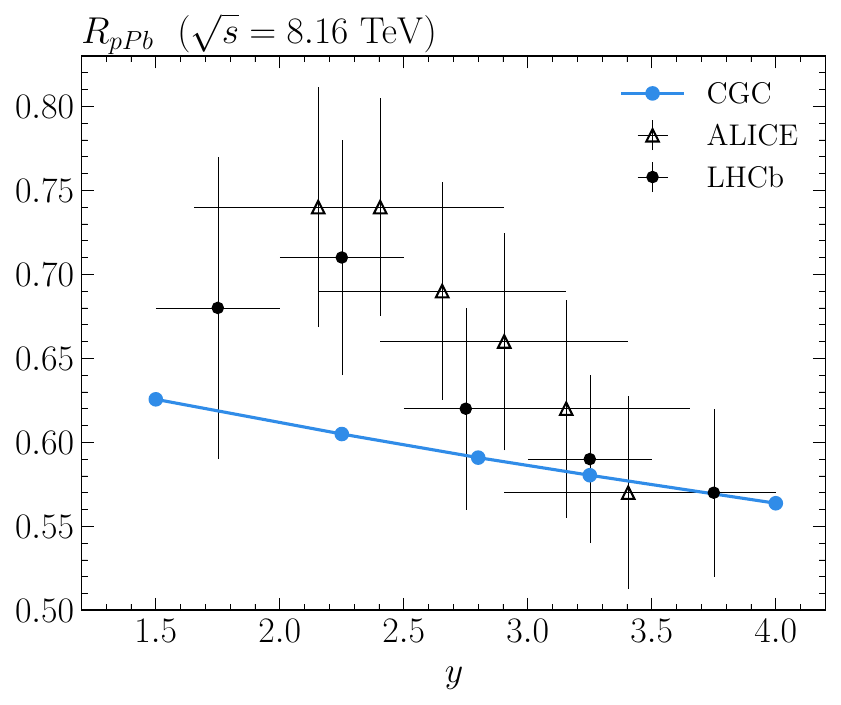}
    \caption{Nuclear supression factor $R_{pPb}$ as a function of rapidity $y$ integrated over the momentum range $0< p_\perp < 10\,\gev$ at  $\sqrt{s}=8.16$ TeV. The vertical error bars are quadratic sums of statistical and systematic uncertainties in both LHCb data~\cite{LHCb:2017ygo} for prompt \jpsi production and ALICE data~\cite{ALICE:2018mml} for inclusive \jpsi production. The horizontal error bars are bin widths.}
    \label{fig: R(Y)}
\end{figure}

In Fig.~\ref{fig: R(Y)} we illustrate the rapidity dependence of the $p_\perp$-integrated $R_{pPb}$ predicted by the BK evolution. The rapidity dependence is weaker than in the data similarly as in Ref.~\cite{Ducloue:2015gfa}, where the color evaporation model was applied with the same CGC setup that is used here. The agreement is better at most forward rapidities, where the dilute-dense factorization is expected to be most accurate.

\section{Summary and discussion}
\label{Conclusions}
In this paper we have calculated the cross section for forward \jpsi production in proton-proton and minimum bias proton-lead collisions in the CGC+NRQCD formalism. 
The CGC description of the dense proton and nuclear target is constrained by HERA DIS data, and the hadronization process is encoded in the LDMEs. Thus there are no free parameters in our setup and we calculate \jpsi production consistently with the DIS data.

We find that the dominant contribution to the cross section is the color octet channel $\prescript{1}{}{S}^{[8]}_0$, which contributes roughly $70\%$ to the cross section in p+p and p+Pb collisions. Interestingly, the color singlet $\prescript{3}{}{S}^{[1]}_1$ contribution is small, with only an approximate $15\%$ contribution to the total. This contrasts with prior predictions based on the Color Singlet Model~\cite{Baier:1981uk} as well as the hierarchy expected from NRQCD power counting in Eq.~\eqref{eq:VectorExpansion}. 

At small $p_\perp$, our predictions for the cross sections calculated as a sum of the individual channels shown in Figs.~\ref{fig: pp cross section} and~\ref{fig: pA cross section} overestimate the experimental LHCb and ALICE data in both p+p and p+Pb collisions. The cross section is also predicted to peak at lower $p_T$ than seen in the data.  We expect that DIS fits including the heavy quark contribution could improve the normalization of the cross section.
Similarly, it could be valuable to revisit the determination of the LDMEs in a consistent fit within the CGC formalism, potentially including polarization observables~\cite{Ma:2018qvc}. Furthermore, moving to NLO can be expected to have a numerically significant effect, for example, the Sudakov factor is expected to push the maximum of the cross section to a larger $p_\perp$.

Unlike absolute cross sections, our predictions for the nuclear modification ratio $R_{pPb}$ as a function of $p_\perp$ shown in Fig.~\ref{fig: R(pt)} are within the error bands of ALICE and LHCb data for most forward rapidites $y\ge2.7$. However, we predict a slightly stronger dependence on the \jpsi transverse momentum than seen in the experimental data.
The  rapidity dependence of $R_{pPb}$ shown in Fig.~\ref{fig: R(Y)} is in agreement with the experimental data at most forward rapidities $y\gtrsim 3$, but a significantly weaker rapidity dependence is predicted closer to midrapidity where the applied dilute-dense factorization is less accurate. This reinforces the validity of the dilute-dense description in the CGC   in the forward region and demonstrates the possibility to use forward inclusive \jpsi data to probe gluon saturation phenomena at LHC energies. 
\begin{acknowledgments}
We thank J. Penttala, F. Salazar and Z. Kang for useful discussions. 
This work was supported by the Research Council of Finland, the Centre of Excellence in Quark Matter (projects 346324 and 364191), and projects 338263, 346567 and 359902, and under the European Union’s Horizon 2020 research and innovation programme by the European Research Council (ERC, grant agreements  No. ERC-2023-101123801 GlueSatLight and ERC-2018-ADG-835105 YoctoLHC) and by the STRONG-2020 project (grant agreement No. 824093). 
Computing resources from CSC – IT Center for Science in Espoo, Finland and the Finnish Grid and Cloud Infrastructure (persistent identifier \texttt{urn:nbn:fi:research-infras-2016072533}) were used in this work.
The content of this article does not reflect the official opinion of the European Union and responsibility for the information and views expressed therein lies entirely with the authors. 

\end{acknowledgments}
\bibliographystyle{JHEP-2modlong}
\bibliography{refs}

\end{document}